\documentclass[draftcls]{IEEEtran}

\usepackage{amsmath,amssymb,amsthm}
\usepackage{bbm}
\usepackage{graphicx,color}

\def\support{\mbox{Rect}[\vec{0}, \vec{1}]}
\def\boundary{{\cal B}[\vec{0}, \vec{1}]}

\title{Mobile Sensing of Two-Dimensional Bandlimited Fields on Random
Paths}

\author{Charvi Rastogi and Animesh Kumar\\
Department of Electrical Engineering \\Indian Institute of Technology Bombay,
Mumbai, India - 400076 \\ Email: rastogicharvi@gmail.com, animesh@ee.iitb.ac.in}
\begin{document}
\maketitle

\begin{abstract}
Mobile sensing has been recently proposed for sampling spatial fields, where 
mobile sensors record the field along various paths for reconstruction.
Classical and contemporary sampling typically assumes that the sampling
locations are approximately known. 
This work explores multiple sampling strategies along random paths to sample and
reconstruct a two dimensional bandlimited field. Extensive simulations are
carried out, with insights from sensing matrices and their properties, to
evaluate the sampling strategies. Their performance is measured by evaluating
the stability of field reconstruction from field samples.  The effect of
location unawareness on some sampling strategies is also evaluated by
simulations.
\end{abstract}

\section{Introduction} 
\label{sec:intro}

Sampling and reconstruction of two dimensional bandlimited fields (signals) is a
well studied subject~\cite{rmarks1,rmarks2}. Recently, mobile sensing was
proposed for sampling spatial fields, where a mobile sensors records the field
along various paths for reconstruction
(interpolation)~\cite{vetterli1,vetterli2,vetterli4}. The sensing paths
considered are mostly deterministic by construction. The main advantage of
mobile sensing is a smaller number of sensing stations, at the cost of its
mobility. Classical sampling and interpolation assumes that the sampling
locations are (approximately) known~\cite{rmarks1,rmarks2}.  Of late, it has
also been shown that a location-unaware mobile sensor can be used to estimate
(reconstruct) a one-dimensional spatially bandlimited field~\cite{akumar2}.

With the above background, this work explores multiple sampling strategies along
random paths to sample and reconstruct a spatially bandlimited field. These
sampling strategies are compared and contrasted in this work. For location
masking, sensors which average out the collected samples are also considered in
some sampling strategies. A two-dimensional bandlimited (finite support) field
has a finite number of non-zero Fourier coefficients. These coefficients represent
the degrees of freedom of the field. One common theme observed is that
oversampling beyond the degree of freedom aids in random path based sampling
strategies. The performance of these sampling strategies is evaluated using the
stability of spatial field reconstruction from field samples.

The main result of this work is the design and evaluation of multiple strategies
for a two-dimensional bandlimited field sampling along random paths. The answers
are obtained by extensive simulations along with intuitive insights.

\textit{Prior art:} Sampling and estimation of bandlimited spatial fields on
equi-spaced parallel paths by mobile sensors is studied and the aliasing error
and measurement-noise is analyzed by Unnikrishnan and Vetterli~\cite{vetterli1}.
Performance of several trajectories for mobile sensing is discussed in
\cite{vetterli2,vetterli4}.  The performance metric used here is path density.
Results from classical sampling theory~\cite{rmarks1,rmarks2} provide schemes
for sampling and estimating the field based on measurements of the field at a
countable number of nonuniform collection of points like the one depicted in
Fig.~\ref{fig:all7models}(a). Location unawareness is introduced in mobile
sensing and distributed sensing for one dimensional field by
Kumar~\cite{akumar2,akumar1}. Observing a spatial field at a random location
is akin to random sensing studied in compressed sensing~\cite{eldar}. 
Tools from compressed sensing, therefore, are useful
in the understanding of random path based sampling schemes.

\section{Preliminaries}
\label{sec:prelim}

\subsection{Field model}
\label{sec:fieldmodel}
It is assumed that $g(x,y)$ is the field of interest. It is two dimensional,
continuous and bandlimited, and supported in $\support := [0,1] \times [0,1]$.
Bandlimitedness implies that 
\begin{align}
g(x,y) = \sum_{k=-b}^{b}\sum_{l=-b}^{b}a[k,l]\exp(j2\pi (kx+ly)),
\label{eq:fourierseries}
\end{align}
where $a[k,l]$ are the Fourier coefficients of $g(x,y)$. The field is assumed to
be temporally fixed (or slowly varying). It is noted that the problem of
estimating a static field with mobile sensor(s) is involved and deserves a first
study. The bandwidth in $(x,y)$ dimensions is assumed to be equal for
simplicity, and does not affect the main results obtained in the paper. The
number of Fourier coefficients is denoted by $n:= (2b+1)^2$, and denotes the
real degrees of freedom. Even though each Fourier coefficient is complex valued,
since $g(x,y)$ is real valued, conjugate symmetry of $a[k,l]$ limits the number
of real degrees of freedom to $n$.  At least $n$ samples of the field are,
therefore, required to reconstruct it.
\subsection{Sampling Schemes} 
\label{sec:samplingintro}

In this work, eight sampling schemes are considered. They are sequentially
described in Section~\ref{sec:samplingmodel}. Largely, two sampling schemes are
considered: point based sampling and path based sampling.  In point based
sampling, samples at various locations are collected in an array.
From~\eqref{eq:fourierseries}, each sample $g(x,y)$ can be expressed as an inner
product between $n$ Fourier coefficients and a complex vector. Thus, $m$ spatial
field samples at $(x_i, y_i), 1 \leq i \leq m$ are given by
\begin{align}
\begin{bmatrix}
g(x_1,y_1) \\
\vdots \\
g(x_m, y_m)
\end{bmatrix} & = 
\begin{bmatrix}
X_{11} & X_{12} &  \dots  & X_{1n} \\
\vdots & \vdots &  \ddots & \vdots \\
X_{m1} & X_{m2} &  \dots  & X_{mn}
\end{bmatrix} 
\begin{bmatrix}
a[-b,-b] \\
\vdots \\
a[b,b]
\end{bmatrix}, \nonumber \\
\mbox{or } \vec{g} & = X \vec{a} \label{eq:gXa}
\end{align}
where $X_{qr} = \exp(j2\pi (k_rx_q+l_ry_q))$. The matrix $X:= [X_{qr}]_{m \times
n}$ will be termed as the sensing matrix. The column parameters $k_r, l_r$
correspond to the Fourier phasors for various $(k,l)$ pairs. The number of
columns is therefore $n$. 

In path based sampling, the mobile sensor is an accumulator which averages all
the measurements made over a path. Averaging is advantageous since it denoises
the readings made, and also does not require location information of individual
samples. Similar to~\eqref{eq:gXa}, a sensing matrix can be formed. If the
sensor samples at $(x_{i,1}, y_{i,1}), \ldots, (x_{i,p_i}, y_{i,p_i})$ on path
$i$ with $p_i$ points, then the sensing matrix $X_{\text{avg}}$ has the
following entries
\begin{align}
X_{\text{avg},ir} = \dfrac{1}{p_i}\Big(\sum_{t=1}^{p_i}\exp\big(j2\pi(k_r
x_{i,t} + l_r y_{i,t})\big)\Big) \label{eq:Xelement}
\end{align}
where $k_r, l_r$ are the values of $k,l$ corresponding to the $r^{\text{th}}$
column, as before.

\begin{figure*}[!htb]
\begin{center}
\includegraphics[width=7.0in]{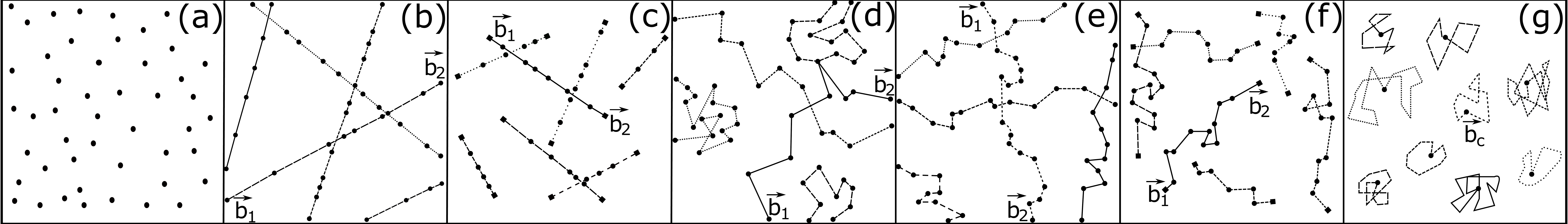}
\end{center}
\caption{\label{fig:all7models} The random paths used for spatial sampling are
illustrated: (a) benchmark sampling scheme where points are uniformly scattered
over the sensing region; (b) the random paths consist of straight lines from
boundary to boundary; (c) the random paths are straight lines between interior
points; (d) the random paths are random walks; (e) the random paths start at end
at designated boundary points; (f) the random paths start and end at designated
interior points; and, (g) the random paths originate and end at randomly
scattered center points.}
\end{figure*}
\subsection{Measurement-noise model} 
In this work, measurement-noise is modeled by an independent and identically
distributed (i.i.d.) process having zero mean and finite variance.  If $W(x_1,
y_1), W(x_2, y_2), \ldots$ are the measurement-noise samples, then they are
i.i.d.~and independent of the field and the random path selected for
sampling.

\section{Sampling Models}
\label{sec:samplingmodel}
Eight different sampling models, used in simulations, are described in this
section. This will help discover randomized sampling schemes from which a
bandlimited field can be estimated, even when sampling locations are unknown and
measurement-noise is present. Most schemes are based on nonuniform random walks
(or simply, random walks).

\subsection{Uniformly scattered fixed location sensors}
\label{sec:scattered}

In this model, static sensors are realized at uniformly distributed locations in
$\support$ (see Fig.~\ref{fig:all7models}(a)).  This static sampling model is
well known in the literature for field reconstruction~\cite{rmarks2}. This model
will give us the benchmark performance of random sampling for comparisons.

%
\subsection{Sampling on random straight line paths}
\label{sec:pointSampling}

Equispaced straight line paths for field sampling were introduced by Unnikrisnan
and Vetterli~\cite{vetterli1}. Let $\boundary$ be the boundary of the region
$\support$.  In this model, a random straight line is chosen by selecting two
independent points $\vec{b}_1$ and $\vec{b}_2$ with the distribution
Uniform$(\boundary)$.  It is assumed that the mobile sensor samples from
$\vec{b}_1$ to $\vec{b}_2$.  The inter-sample spacings are chosen according to
$D \sim \mbox{Uniform}(0, \gamma)$ distribution, where $\gamma > 0$ controls the
average sample spacing on the path (see Fig.~\ref{fig:all7models}(b)). With path
angle $\theta$ with respect to $x$-axis, the $t$-th field sample on a path is
observed at 
\begin{align}
x_{t+1} =  x_t + d_t\cos(\theta), \quad y_{t+1} =  y_t + d_t\sin(\theta).
\label{eq:randwalk}
\end{align}
If $p$ paths and an average $\Gamma$ samples/path are selected for sampling,
then the sensing matrix in~\eqref{eq:gXa} will be of the size $\Gamma p \times
n$. To avoid an underdetermined system in~\eqref{eq:gXa}, $p > 2b+1$ and $\Gamma
\geq 2b + 1$ will be selected.

\subsection{Averaging over random straight line paths}
\label{sec:boundaryStraight}

In this sampling model, a path and sampling locations are selected as in
Section~\ref{sec:pointSampling}. However, all the samples along the path are
averaged out by the mobile sensor to conserve storage and mask sampling
locations. If $m$ is the number of paths, then the sensing matrix
in~\eqref{eq:Xelement} is of size $m \times n$.

\subsection{Straight line path between two inner points} 
\label{sec:innerStraight}

In this sampling model, a random straight line is chosen by selecting two
independent points $\vec{b}_1$ and $\vec{b}_2$ according to a Uniform$(\support)$
distribution. The sensor averages samples over a path and traverses according to
the rule in \eqref{eq:randwalk} (see Fig.~\ref{fig:all7models}(c)). If $m$ is
the number of paths, the sensing matrix is of size $m \times n$ as
in~\eqref{eq:Xelement}.

\subsection{Random walk}
\label{sec:randomWalk}

In this model, the mobile sensor starts at a point $\vec{b}_1$ chosen uniformly
in $\boundary$. Then, the sensor traverses at each step using a random
step-size $D \sim \text{Uniform}(0, \gamma)$ and an angle $\theta \sim
\text{Uniform}(0, 2\pi)$. Sampling locations on the path are given
by~\eqref{eq:randwalk} with $\theta_t$ instead of $\theta$ (see
Fig.~\ref{fig:all7models}(d)).  In this model, the sensor may exit the boundary
close to $\vec{b}_1$. For small $\gamma$, the sensor may sample at a 
large number of points as well.

\subsection{Directed random walk between boundary points}
\label{sec:boundaryRandomWalk}

In this model, two independent points $\vec{b}_1$ and $\vec{b}_2$ are selected
according to a Uniform$(\boundary)$ distribution. Path with end points on the
same edge of the boundary are rejected. A random walk with $p$ steps is used to
create a directed random walk from $\vec{b}_1$ to $\vec{b}_2$. The $p$-step
random walk is implemented according to~\eqref{eq:randwalk} and $(x_1,y_1) =
\vec{b}_1$. Then to ensure the random walk is directed, the points are modified
as
\begin{align}
(x'_t,y'_t) = (x_t , y_t) + \frac{t}{p}(\vec{b}_{2} - (x_p,y_p))
\label{eq:brownian}
\end{align} 
with $1 \leq t \leq p$. Note that $(x'_1,y'_1) = \vec{b}_1$ and $(x'_p,y'_p) =
\vec{b}_2$.  These paths are illustrated in Fig.~\ref{fig:all7models}(e) and are
inspired from Brownian motion and Brownian bridge~\cite{durrett}.

\subsection{Directed random walk between two inner points}
\label{sec:innerRandomWalk}

In this model, two independent points, $\vec{b}_1$ and $\vec{b}_2$ distributed
according to Uniform$(\support)$ are generated. A directed random walk using the
setup in~\eqref{eq:brownian} is generated (see Fig.~\ref{fig:all7models}(f)).

\subsection{Bee and hive sampling model}
\label{sec:beehive}

In this model, a point $\vec{b}_c$ is selected according to Uniform$(\support)$
distribution. The mobile sensor starts and ends its random walk at $\vec{b}_c$
using the setup in~\eqref{eq:brownian}.  A total of $m$ such points are
generated. This sampling model is inspired by bees that hover around their hive
and then return to the same location. The sensor is assumed to be location
unaware. This sensing matrix will be of size $m \times n$.

\section{Spatial field estimation method}
\label{sec:estimation}

The general method of reconstruction is explained. Since linear measurements are
obtained, a regression based reconstruction is natural.  From the samples
collected by the mobile sensors (as described in
Section~\ref{sec:samplingmodel}), a regression style estimate of Fourier
coefficients $a[k,l]$ are obtained as follows:
\begin{align}
\widehat{\vec{a}}  = (X^* X)^{-1} X^* \vec{g} := {\cal Y} \vec{g}
\label{eq:estimation}
\end{align}
where $\vec{g}$ are the field samples obtained either by taking samples or their
averages along a path, and $X$ is the sensing matrix (see~\eqref{eq:gXa}).

If the sensor is location unaware, the sensing matrix is $X_{\text{un}}$, which
is formed by approximating the locations using the end-points $\vec{b}_1,
\vec{b}_2$ of the path and the number $p$ of measurements made. The locations
are approximated as 
\begin{align}
(x_{t}, y_{t})_{\text{un}} = \vec{b}_1 + \dfrac{t}{p-1}(\vec{b}_2 - \vec{b}_1),
\quad t = 0, 1, \ldots, p-1.\label{eq:location_approx}
\end{align}
See Section~\ref{sec:results} for the effect of location unawareness. 

To test the feasibility of sampling, the stability of pseudo inverse
in~\eqref{eq:estimation} has to be characterized. Analysis of the stability is
very difficult for all the sampling schemes  presented. So, we will adopt a
simulation based approach. The stability of pseudo-inverse will be quantified
using the condition number $C$ of a matrix. If samples are quantized or affected
by independent measurement-noise, then a small condition number ensures that the
estimate $\widehat{\vec{a}}$ is not too noisy. An ill-posed problem gives an
unstable inverse and has a very high condition number.  The condition number is
defined as~\cite{eldar}
\begin{align}
C_2 (X) = \frac{\sigma_{\text{max}}(X)}{\sigma_{\text{min}}(X)} =
\Big(\frac{\lambda_{\text{max}}(X^*X)}{\lambda_{\text{min}}(X^*X)}\Big)^{\frac{1}{2}}
\label{eq:condition2}
\end{align}
where $\sigma$ and $\lambda$ denote the singular and eigenvalues, respectively.
The condition number in our simulations depends on number of samples/paths $m$,
the step-size parameter $\gamma$, and the location awareness/unawareness of
sensor. The results are discussed next.

\begin{figure*}[!htb] 
\begin{center}
\includegraphics[width=7.0in]{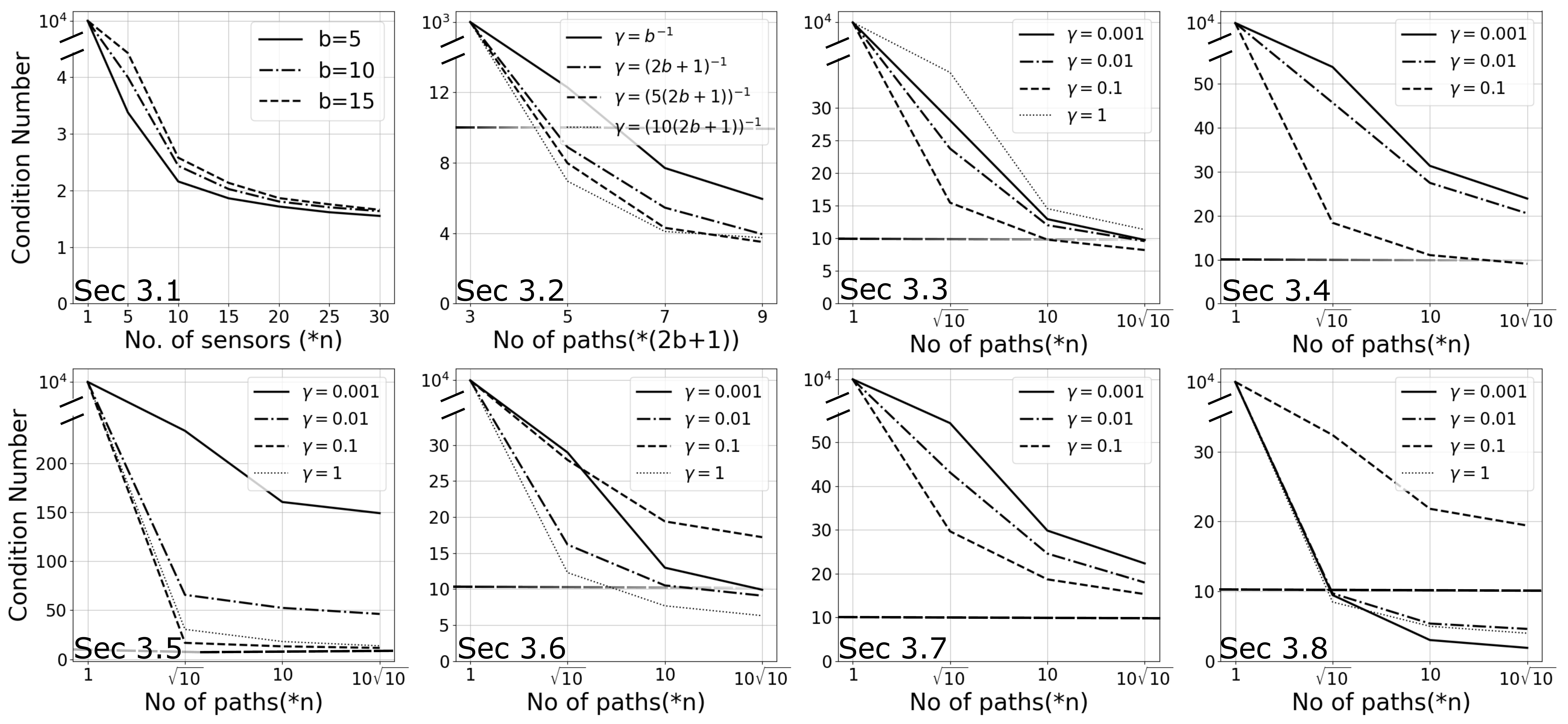}
\end{center}
\caption{\label{fig:condition} Condition number plots for various strategies are
illustrated. The parameters $m$ and step-size $\gamma$ are varied.  For
uniformly scattered non mobile sensors, the variation is shown for different
values of bandwidth parameter, $b$, and for the rest $b=10$ is fixed.  These
results are obtained by averaging over 50 iterations. The smallest condition
numbers are achieved by schemes of Sec.~3.2 and Sec.~3.8.}
\end{figure*}
\section{Results}
\label{sec:results}

The condition number corresponding to various sampling models in
Section~\ref{sec:samplingmodel} are presented. These results are obtained by
averaging $C_2(X)$ over 50~iterations. For computational efficiency in
determining $C_2(X)$, the eigenvalues of $X^*X$ are calculated as it is a
smaller matrix than $X$ (see~\eqref{eq:condition2}). As noted earlier, a low
condition number is more desirable. The number of rows $m$ in the sensing matrix
$X$ is always taken to be $\geq n = (2b+1)^2$ since it is necessary for the
pseudo-inverse of $X$ to exist. So, the number of samples (or paths) is a
multiple of $(2b+1)^2$ along the $x$-axis. All the results are illustrated in
Fig.~\ref{fig:condition} and the plots are explained next.

A common trend is that with increasing size of sensing matrix $m$ the condition
number improves. The benchmark sampling method in Sec.~\ref{sec:scattered}
achieves a condition number lower than $10$ for moderate values of $m$. The
benchmark is the best performance to hope for, in terms of condition number. The
results for sampling models in Sec.~\ref{sec:pointSampling} and
Sec.~\ref{sec:beehive} are the best among random sampling strategies and near to
the benchmark performances. This is especially surprising for
Sec.~\ref{sec:beehive} where the location information of sensor is also unknown.
The schemes of Sec.~\ref{sec:boundaryStraight} and
Sec.~\ref{sec:boundaryRandomWalk} which operate with paths starting and ending
at the boundary are fair; while, their counterparts in
Sec.~\ref{sec:innerStraight} and Sec.~\ref{sec:innerRandomWalk} with interior
boundary points are average in performance. The worst scheme is of
Sec.~\ref{sec:randomWalk}, which consists of a fully random walk.

The condition number trends for all the schemes except
Sec.~\ref{sec:pointSampling} can be explained using condition number results
from random matrix theory. From~\cite{eldar}, it is known that if $X$ is an $m
\times n$ sensing matrix with independent and sub-Gaussian rows, then for every
$t > 0$ with probability $\geq 1 - e^{(-ht^2)}$
\begin{align}
1 \leq \frac{\sigma_{\text{max}}(X)}{\sigma_{\text{min}}(X)} = C_2(X) \le
\frac{\sqrt{m} + H\sqrt{n} + t}{\sqrt{m} - H\sqrt{n} - t}, 
\label{eq:condNumberBound}
\end{align}
where $h,H >0$ are finite constants that depend only on the sampling model and
not on $(n, m)$. So, condition number $C_2(X)$ provably improves with $m$.

As step-size $\gamma$ decreases, \textit{location unawareness} of mobile sensors
can also be introduced. Loosely speaking, as the average sampling rate over a
straight line path increases, a random distribution of the points on the path
averages to the equispaced points~\cite{akumar2}.  Therefore, the sensing matrix
$X_{\text{un}}$ can be approximated as in \eqref{eq:location_approx}.  This is
applicable to schemes in Sec.~\ref{sec:pointSampling},
Sec.~\ref{sec:boundaryStraight}, and Sec.~\ref{sec:innerStraight}. In scheme of
Sec.~\ref{sec:beehive}, location-unawareness works since spatial field's
variation gets averaged out over the random walk for small step-size $\gamma$.

Finally, as the number of rows of sensing matrix $X$ increases,
\textit{measurement-noise filtering} naturally happens by regression
in~\eqref{eq:estimation}. A small condition number also ensures that
measurement-noise power is less amplified~\cite{bickelDM2001}.

\section{Conclusions}

Multiple sampling strategies along random paths to sample and reconstruct a two
dimensional bandlimited field were explored. Using simulations it was found that
a bee and hive based location-unaware random sampling design has the best
condition number among various random sampling strategies. A close second is random
straight-lines based sampling strategy. Most of the obtained results can be
explained by using the condition number results from random matrix theory.

\bibliographystyle{IEEEtran}
\bibliography{refs}
\end{document}